\documentstyle[12pt]{article}
\input psfig.tex
\begin{document}

\title{Initial Condition Sensitivity of Global Quantities in 
      Magnetohydrodynamic Turbulence}
\author{Gaurav Dar\thanks{e-mail: gdar@iitk.ernet.in}, 
Mahendra K. Verma \\ 
Department of Physics  \and V. Eswaran \\ 
Department of Mechanical Engineering \\ Indian Institute of Technology, 
Kanpur 208016, India}
\maketitle

% ....................... ABSTRACT ................................

\begin{abstract}
In this paper we study the effect of subtle changes in initial conditions
on the evolution of global quantities in two-dimensional Magnetohydrodynamic
(MHD) turbulence. 
We find that a change in the initial phases of complex Fourier modes of the 
Els\"{a}sser variables, while keeping the initial values of total energy, 
cross helicity and Alfv\'{e}n ratio unchanged, has a significant effect on 
the evolution of cross helicity. On the contrary, the total energy and
Alfv\'{e}n ratio are insensitive to the initial phases. Our simulations are
based on direct numerical simulation using the pseudo-spectral method.
\end{abstract}

PACS Number(s): 47.65.+a,47.27.Gs,47.27.Jv 

\bibliographystyle{unsrt}

\newpage
\topmargin -0.5in
\textheight 9.0in

%........................ THE DETAILS ................................
In  fluid turbulence, the evolution of the velocity, ${\bf u}({\bf x})$,
at a given position or of a given Fourier component, ${\bf u}({\bf k})$,
is known to be sensitive to the details of the initial conditions, e.g. phases
of ${\bf u}({\bf k})$ (to be defined rigorously later). However, the evolution
of global quantities like total energy are generally presumed to depend 
only on initial values of total energy and energy spectrum. The averaging
over many modes appear to wash out the effects of the initial phases after a
reasonably long time. The total energy in simulations with the same initial 
energy and spectrum, but with the modes chosen randomly, evolve along 
nearby trajectories; This is demonstrated in Fig. 1. It has been a common 
belief that in MHD turbulence also the evolution of global quantities, 
e.g. total energy and cross helicity, depends only on the initial values of 
global quantities and their spectra. The evolution of global quantities have 
earlier been studied by Ting et al. \cite{Ting86}, Matthaeus et al. 
\cite{Matth184,Matth284}, Biskamp and welter \cite{Bisk89}, Pouquet et al. 
\cite{Pouq88} in which they found dynamic alignment and various other phenomena. 
In this paper we show numerically that under certain conditions in 
two-dimensional (2D) magnetohydrodynamic (MHD) turbulence, the evolution of 
global quantities may not depend solely only on their initial values, but may 
depend significantly on more subtle features like the phases of complex Fourier
modes (to be defined below) of the initially prescribed fields of the dynamical 
variables. In other words, a knowledge of the gross initial features as 
specified by the global quantities is not sufficient under all conditions to 
determine the evolution of global quantities. We only choose phases as a 
convenient way of demonstrating the inadequacy of specifying the initial 
global quantities and spectra alone for computing the evolution of certain 
global quantities. We find that the evolution of cross helicity shows a 
sensitivity to the initial phases in simulations with  small values of initial
cross helicity.  

The primary variables in MHD turbulence are the velocity field ${\bf u}$ and 
the magnetic field ${\bf b}$. In our simulations we take mean magnetic field
to be zero. We use Els\"{a}sser variables ${\bf z}^{\pm }={\bf u \pm b}$ in our simulations. 
These variables denote fluctuations with 'positive' and 'negative' 
velocity-magnetic field correlations. The relevant quadratic quantities are  %
\begin{eqnarray}
E^{+} & = & \frac{1}{2} \int_{unit \: vol.} \left({\bf z}^{+}\right)^{2}
d{\bf v}  = \frac{1}{2} \sum_{{\bf k}} \left( \left| {\bf z}^{+}
({\bf k}) \right|^{2} \right),       \label{eq:zpdef}
\end{eqnarray}
\begin{eqnarray}
E^{-} & = & \frac{1}{2}\int_{unit \: vol} \left({\bf z}^{-}\right)^{2}% 
d{\bf v} = \frac{1}{2} \sum_{{\bf k}} \left( \left| {\bf z}^{-}({\bf k})%
 \right|^{2} \right),      \label{eq:zmdef}
\end{eqnarray}
the magnetic energy,
\begin{eqnarray}
E_{b} & = & \frac{1}{2}\int_{unit \: vol.} {\bf b}^{2} d{\bf v}  \nonumber \\
      & = & \frac{1}{8} \sum_{{\bf k}} \left( \left| {\bf z}^{+}({\bf k})%
\right|^{2} + \left| {\bf z}^{-}({\bf k}) \right|^{2} - 2  Re \left({\bf z}^{+}({\bf k}) \cdot %
{\bf \tilde z}^{-}({\bf k})\right) \right),   \label{eq:bdef} 
\end{eqnarray}
the fluid energy,
\begin{eqnarray}
E_{u} & = & \frac{1}{2}\int_{unit \: vol.} {\bf u}^{2} d{\bf v}  \nonumber \\
      & = & \frac{1}{8} \sum_{{\bf k}} \left( \left| {\bf z}^{+}({\bf k}) 
\right|^{2} + \left| {\bf z}^{-}({\bf k}) \right|^{2} + 2  Re \left({\bf z}^{+}({\bf k}) \cdot %
{\bf \tilde z}^{-}({\bf k})\right) \right).     \label{eq:udef} 
\end{eqnarray} 
and the mean square magnetic vector potential
\begin{eqnarray}
A & = & \sum_{\bf k} \left |\psi({\bf k}) \right |^{2} \nonumber \\
  & = & \sum_{\bf k} \frac {\left |{\bf b}({\bf k}) \right |^{2}}{k^{2}}  \label{eq:vecpdef}
\end{eqnarray}

\noindent where $\psi$ is the magnetic vector potential, ${\bf \tilde z}^{+}$ 
and  ${\bf \tilde z}^{-}$ are complex conjugates of ${\bf z}^{+}$ and 
${\bf z}^{-}$ respectively. The total energy is $E=(E^{+}+E^{-})/2$ and the 
cross helicity is $H_{c}=(E^{+}-E^{-})/2$. There are two well known 
dimensionless parameters: the normalised cross helicity $\sigma_{c}=H_{c}/E$ 
and the Alfv\'{e}n ratio $r_{A}=E_{u}/E_{b}$. The total energy, cross helicity,
and the mean square vector potential are the three global inviscid invariants 
of the 2D MHD equation.

We denote the complex Fourier modes ${\bf z}^{\pm }({\bf k})$ by $\left|%
{\bf z}^{\pm}({\bf k})\right|\exp (i\theta _{{\bf k}}^{\pm })$, 
where $\theta_{{\bf k}}^{\pm}$ are the phases of the modes. All the three
global inviscid invariants $E$, $H_{c}$, and $A$  are independent of phases,
while the  Alfv\'{e}n ratio $r_A$  depends on the phase difference 
$\theta_{\bf k}^{+}-\theta_{\bf k}^{-}$. Ting et al. \cite{Ting86} found that
the Alfv\'{e}n ratio affects the evolution of global quantities; it follows 
from their observations that the  initial phase difference 
$\theta_{{\bf k}}^{+}-\theta_{{\bf k}}^{-}$ would affect the global evolution.
In this paper we demonstrate numerically that even keeping this initial phase 
difference fixed, change of absolute value of the initial phases  
$\theta^{+}_{\bf k}$ affects the global evolution.

In our simulations we investigate the effects of the initial phases on the 
subsequent total energy $E$, normalised cross helicity $\sigma_c$, and 
Alfv\'{e}n ratio $r_A$. The temporal evolution of $\sigma_c$ has been the 
subject of investigation in a number of earlier studies \cite{Ting86,Matth184,
Matth284,Bisk89,Pouq88}. In several of these studies, $\sigma_c$ has been 
observed to increase with time \cite{Matth184,Matth284,Bisk89,Pouq88}, a 
behaviour termed as dynamic alignment.  However, Biskamp and Welter%
\cite{Bisk89} observed in their simulations that the tendency towards 
dynamic alignment decreases with the increase in Reynolds number, and  
$\sigma_c$ could even decrease at high enough Reynolds number \cite{Bisk89}. 
Ting et al. \cite{Ting86} too observed a few cases of decreasing $\sigma_c$ 
for small values of initial $\sigma_c$ and $E/A$. In these earlier studies
the effects of absolute phases have not been studied. 

We solve the 2-D incompressible MHD equations with hyperviscosity.  The 
equations written in terms of the Els\"{a}sser variables, ${\bf z}^{+}$ and 
${\bf z}^{-}$ are
\begin{eqnarray}
\frac{\partial{\bf z}^{\pm}}{\partial{t}} \mp 
\left({\bf B}_{0}.{\bf \nabla}\right) {\bf z}^{\pm} +
\left ({\bf z}^{\mp}.{\bf \nabla}\right){\bf z}^{\pm} & = & -{\bf \nabla}p +
 \nu_{\pm} \nabla^{2}{\bf z}^{\pm} + 
\nu_{\mp} \nabla^{2}{\bf z}^{\mp} \nonumber \\
 &  & + \left(\nu_{\pm}/k_{eq}^{2}\right)\nabla^{4}{\bf z}^{\pm} 
      + \left(\nu_{\mp}/k_{eq}^{2}\right)\nabla^{4}{\bf z}^{\mp} \label{eq:mhd}
\end{eqnarray}
where $\nu_{+}$ and $\nu_{-}$ are related to the fluid viscosity $(\nu )$
and magnetic diffusivity $(\mu )$ by the relationship $\nu ^{\pm }=1/2(\nu \pm
\mu )$. The last two terms in Eq. (6) include hyperviscosity  
$\nu^{\pm}/k^{2}_{eq}$ to damp out the energy at very high wave numbers.
We choose $\nu = \mu = 5 \times 10^{-4}$ for runs on a grid of size 
$512 \times 512$ and $\nu = \mu = 10^{-3}$ for runs on a grid of size 
$256 \times 256$. The hyperviscosity related parameter $k{\rm_{eq}}$ is chosen
to be 20 for runs on both the grids.  The time step $dt$ used for these runs 
is $10^{-3}$. The simulations are carried up to the final time $t_{final}=50$.

We use the pseudo-spectral method \cite{Orszag72,Eswar188,MKV96} to solve 
the above equations in a periodic box of size $2\pi \times 2\pi $. In order 
to remove the aliasing errors arising in the pseudo-spectral method a 
square truncation is performed wherein all modes with $|k_x|\geq N/3$ or 
$|k_y|\geq N/3$ are set equal to zero. The equations are time advanced 
using the second order Adam-Bashforth scheme for the convective terms and 
Crank-Nicholson for the viscous terms. In order to validate our code we used
a simulation result of Pouquet et al. \cite{Pouq88} for comparison (see Fig. 2).

The simulations are performed for various initial sets of $\sigma _c$ and 
$r_A$ values. The initial conditions are generated by first fixing $r_{A}$, 
$\sigma_{c}$, $E^{+}$, and $E^{-}$. The chosen value of $r_A$ determines the 
phase difference $\theta_{\bf k}^{+}-\theta_{\bf k}^{-}$.  The initial $E$ 
and $\sigma _c$ determine $|{\bf z}^{\pm}({\bf k})|$.  Note that the absolute
phase $\theta^{+}_{\bf k}$ is still a free parameter. Only modes within the 
annular region $1/2\leq |{\bf k}|<3/2$ are non-zero and each of the modes 
within this region receives equal energy 
(i.e., $|{\bf z}^{\pm }({\bf k})|^2=E^{\pm }/M$, where M is the number of 
modes in the shell). The initial states are generated thus for only half the 
modes - the remaining half are conjugate to them. 

The phase sensitivity of the evolution of $\sigma_c$, E, and $r_A$ are studied
by comparison of pairs of simulations in which initial $\theta^{+}_{\bf k}$ 
are different. We change the initial phases in two ways. In one case we change
$\theta^{+} _{\bf k}$ uniformly for all the modes by an amount $\Delta$, 
while in the other case the phases are changed by using a different random 
seed in the random number generator.  The initial global quantities E, $H_c$, 
$r_A$, and their spectra remain unchanged under these phase changes. The 
evolution of $\sigma_c$ for a variety of initial $\sigma_c$, $r_A$ and 
$\Delta$ values are shown in Table 1 (pairs of simulations are shown together; 
for example, mhd1 differs from mhd1$^*$ only in that its initial 
${\bf z}^{\pm}({\bf k})$ fields have to be shifted from the latter's by 
$\Delta=0.4$). 

The $N=512$ runs with $t_{final}=50$ are very time intensive.  Hence only the
small initial $\sigma_c$ runs, which we found to be sensitive to the phases, 
were carried out for N=512. A large number of runs were performed on N=64 to
explore a wider range of initial conditions. All these results showed
behaviour consistent with the results discussed below which are based on
the high resolution runs N=256 and 512. 

In our simulations, the small $\sigma_c$ runs showed the most significant 
dependence on initial phase $\theta^{+}_{\bf k}$.  For $\sigma_c=0.1$ and 
$r_A=1.5$, we choose $\Delta$ to be 0.0 and 0.4 in mhd1 and mhd1$^*$ 
respectively. It is seen in Fig. 3 that phase shifting has a marked effect on 
the evolution of $\sigma_c$.  For $\Delta = 0.0$ (mhd1) $\sigma_c$ increases 
from its initial value of 0.1 to its final value of 0.129, whereas for 
$\Delta=0.4$ (mhd1$^*$), $\sigma_c$ decreases to a final value of 0.06. The 
total energy (Fig. 4) and the Alfv\'{e}n ratio (Fig. 5) do not appear to be 
affected much by the phase shift. We also  compare two simulations (mhd1 and 
mhd1$^{**}$ in Table 1.) in which the initial phases are generated using 
different random seeds. For these cases also  the effect on the evolution of 
$\sigma_c$ (Fig. 3) is significant, but the corresponding effects on the 
evolution of the total energy (Fig. 4) and $r_A$ (Fig. 5) are not noticeable. 
We also studied the effects of initial phases for the same initial $\sigma_c$,
but with a large initial $r_A (5.0)$. The runs mhd2 and mhd2$^*$ show the 
effects of changing $\Delta$, while mhd2 and mhd2$^{**}$ show the effects of 
different random number generator seeds. The results obtained for this case are 
similar to the run for initial condition with $r_A=1.5$. In Fig. 6 it is seen 
that for initial value of $\Delta=0.0$ (mhd2), $\sigma_c$ increases and for 
initial $\Delta=0.3$, $\sigma_c$ decreases. The effect of changing $\Delta$ on 
the total energy (Fig. 7) and $r_A$ (Fig. 8) is seen to be small. Similar 
results are obtained if we change the seed of the random number generator 
(compare mhd2 and mhd2$^{**}$). Hence, $\sigma_c$ (Fig. 6) is sensitive to the
change in the initial phases whereas the total energy (Fig. 7) and $r_A$ 
(Fig. 8) are not sensitive. Earlier, Ting et al. \cite{Ting86} had observed a 
decrease in $\sigma_c$ for small initial $\sigma_c$. 

We also perform runs at higher initial values of $\sigma_c$ (mhd3 and 
mhd3$^*$ in Table 1.). The effect of phase shifting for initial 
$\sigma_c=0.5$ and $r_A=5.0$ is shown in Figures 9,10,11. It is seen in Fig. 9
that the changes in evolution caused by phase shifting are relatively smaller
for high initial $\sigma_c$ values as compared to small initial $\sigma_c$ 
discussed above. From Fig. 10 it can be seen that the effect of $\Delta$ on 
total energy  is also small and $r_A$ (Fig. 11) remains insensitive to the 
change in $\Delta$. We have performed more runs than have been shown here and 
in all cases $\sigma_c$ was seen to increase for high $\sigma_c$ values. This
increase in $\sigma_c$ is consistent with simulations performed earlier
for high $\sigma_c$ values \cite{Ting86,Matth184,Matth284,Bisk89,Pouq88}.

From the numerical results presented here we conclude that phases of the
initial modes play an important role in the evolution of $\sigma_c$,
atleast for cases with small initial $\sigma_c$ values. For higher values of 
$\sigma_c$, phases do not appear to affect the evolution of $\sigma_c$ by 
any significant amount. In all the runs the total energy and $r_A$ were 
seen not to have any significant dependence on the phases. 

The origin of the phase effects discussed here is not clear at this moment.
We need to examine the evolution more carefully before reaching any definite 
conclusion.  These studies could find applications in understanding the solar
wind observations in which $\sigma_{c}$ has been observed to decrease 
\cite{Robert87a,Robert87b,Marsch90,Matth82a}.

It has been demonstrated in the paper that the evolution of normalised 
cross helicity is significantly affected by subtle features of the initial 
condition especially at low initial cross helicities. This observation will
require us to be more circumspect in drawing conclusions based on arbitrary 
initial conditions and to exercise more care in choosing the initial conditions
in MHD turbulence.

We are thankful to  Dr. R. K. Ghosh for providing us computer time through 
the project TAPTEC/COMPUTER/504 sponsored by All India Council for Technical
Education (AICTE). This work was also supported by Department of Science
and Technology (DST) India project SR/SY.P-11/94.

\newpage

% ........................... REFERENCES ............................

\newpage

\begin{center}
{\bf FIGURE CAPTIONS}
\end{center}

\vspace{2.0cm}

\noindent Figure 1. Energy evolution for fluids. The two cases shown here 
differ only in the phases of the initial conditions. The modes in the runs were
$|{\bf u}({\bf k})|\exp^{i(\theta_{\bf k}+\Delta)}$ with $\Delta=0.0$ and
$0.4$.

\vspace{1.0cm}

\noindent Figure 2. Time evolution of kinetic (solid line) and magnetic 
(dashed line) energies for the O-T vortex using a set of parameters from 
Pouquet et al. (Pouquet et al's results are shown by diamonds and crosses).  
Here $\nu = 2.5 \times 10^{-3}$  and  hyperviscosity is zero.

\vspace{1.0cm}

\noindent Figure 3. Normalised cross helicity ($\sigma_{c}$) evolution for
initial $\sigma_{c}=0.1$, $r_{A}=1.5$. The curves shown correspond to
mhd1, mhd1$^*$, and mhd1$^{**}$ in Table 1.

\vspace{1.0cm}

\noindent Figure 4. Evolution of total energy ($E$) for same initial
conditions as in Fig. 3. See Table 1. for description of mhd1, mhd1$^*$,
and mhd1$^{**}$.

\vspace{1.0cm}

\noindent Figure 5. Evolution of Alfv\'{e}n ratio ($r_A$) for same initial 
conditions as in Fig. 3. Look at Table 1. for description of mhd1, mhd1$^*$,
and mhd1$^{**}$.

\vspace{1.0cm}

\noindent Figure 6. Normalised cross helicity ($\sigma_{c}$) evolution for
initial $\sigma_{c}=0.1$, $r_{A}=5.0$. The curves shown correspond to
mhd2, mhd$2^*$, and mhd2$^{**}$ in Table 1.

\vspace{1.0cm}

\noindent Figure 7. Evolution of total energy ($E$) for same initial
conditions as in Fig. 6. Look at Table 1. for description of mhd2, mhd$2^*$,
and mhd2$^{**}$.

\vspace{1.0cm}

\noindent Figure 8. Evolution of Alfv\'{e}n ratio ($r_A$) for same initial 
conditions as in Fig. 6. Look at Table 1. for description of mhd2, mhd2$^*$,
and mhd2$^{**}$.

\vspace{1.0cm}

\noindent Figure 9. Normalised cross helicity ($\sigma_{c}$) evolution for
initial $\sigma_{c}=0.5$ and $r_{A}=5.0$. The curves shown correspond to
mhd3, mhd3$^*$ in Table 1.

\vspace{1.0cm}

\noindent Figure 10. Evolution of total energy ($E$) for same initial
conditions  as in Fig. 9. Look at Table 1. for description of mhd3, mhd3$^*$.

\vspace{1.0cm}

\noindent Figure 11. Evolution of Alfv\'{e}n ratio for same initial conditions
as in Fig. 9. Look at Table 1. for description of mhd3, mhd3$^*$.

\newpage

% .........................  THE TABLE ..........................

\begin{table}
\caption{Initial values of the random number generator seed $\Delta$,
$\sigma_c$ and $r_A$ for runs performed on grid of size $N \times N$. The
initial and the final values (at $t_{final}=50$) of $\sigma_c$ are also
shown.}
\begin{tabular}{cccccccc} \hline \hline
Run  &  N  & Seed & $r_A$ &  $\Delta$  &  $\sigma_{c}(t=0)$ & $\sigma_{c}(t=50)$ &  $\sigma_{c}$ increases/decreases  \\ \hline
mhd1           & 512 & 50  & 1.5 & 0.  &  0.1 & 0.06   & decreases \\
mhd1$^*$       & 512 & 50  & 1.5 & 0.4 &  0.1 & 0.13   & increases \\
mhd1$^{**}$    & 512 & 575 & 1.5 & 0.  &  0.1 & 0.20   & increases  \\ 
mhd2           & 512 & 50  & 5.0 & 0.  &  0.1 & 0.22   & increases \\
mhd2$^*$       & 512 & 50  & 5.0 & 0.3 &  0.1 & 0.05   & decreases \\
mhd2$^{**}$    & 512 & 575 & 5.0 & 0.  &  0.1 & 0.13   & increases  \\ 
mhd3           & 256 & 50  & 5.0 & 0.  &  0.5 & 0.87   & increases \\
mhd3$^*$       & 256 & 50  & 5.0 & 0.4 &  0.5 & 0.88   & increases \\ 
mhd4           & 64  & 50  & 1.5 & 0.  &  0.1 & 0.02   & decreases \\
mhd4$^*$       & 64  & 50  & 1.5 & 0.6 &  0.1 & 0.34   & increases \\
mhd4$^{**}$    & 64  & 50  & 1.5 & 1.0 &  0.1 & 0.17   & increases \\ 
mhd5           & 64  & 50  & 2.0 & 0.  &  0.1 & - 0.02 & decreases \\
mhd5$^*$       & 64  & 50  & 2.0 & 0.2 &  0.1 & - 0.01 & decreases \\
mhd5$^{**}$    & 64  & 50  & 2.0 & 0.4 &  0.1 & 0.22   & increases \\ 
mhd6           & 64  & 50  & 1.0 & 0.  &  0.5 & 0.79   & increases \\
mhd6$^*$       & 64  & 50  & 1.0 & 0.2 &  0.5 & 0.74   & increases \\ 
mhd6$^{**}$    & 64  & 50  & 1.0 & 0.4 &  0.5 & 0.72   & increases \\ \hline \hline
\end{tabular}
\end{table}

\clearpage

\begin{figure}[p]
\centerline{\mbox{\psfig{file=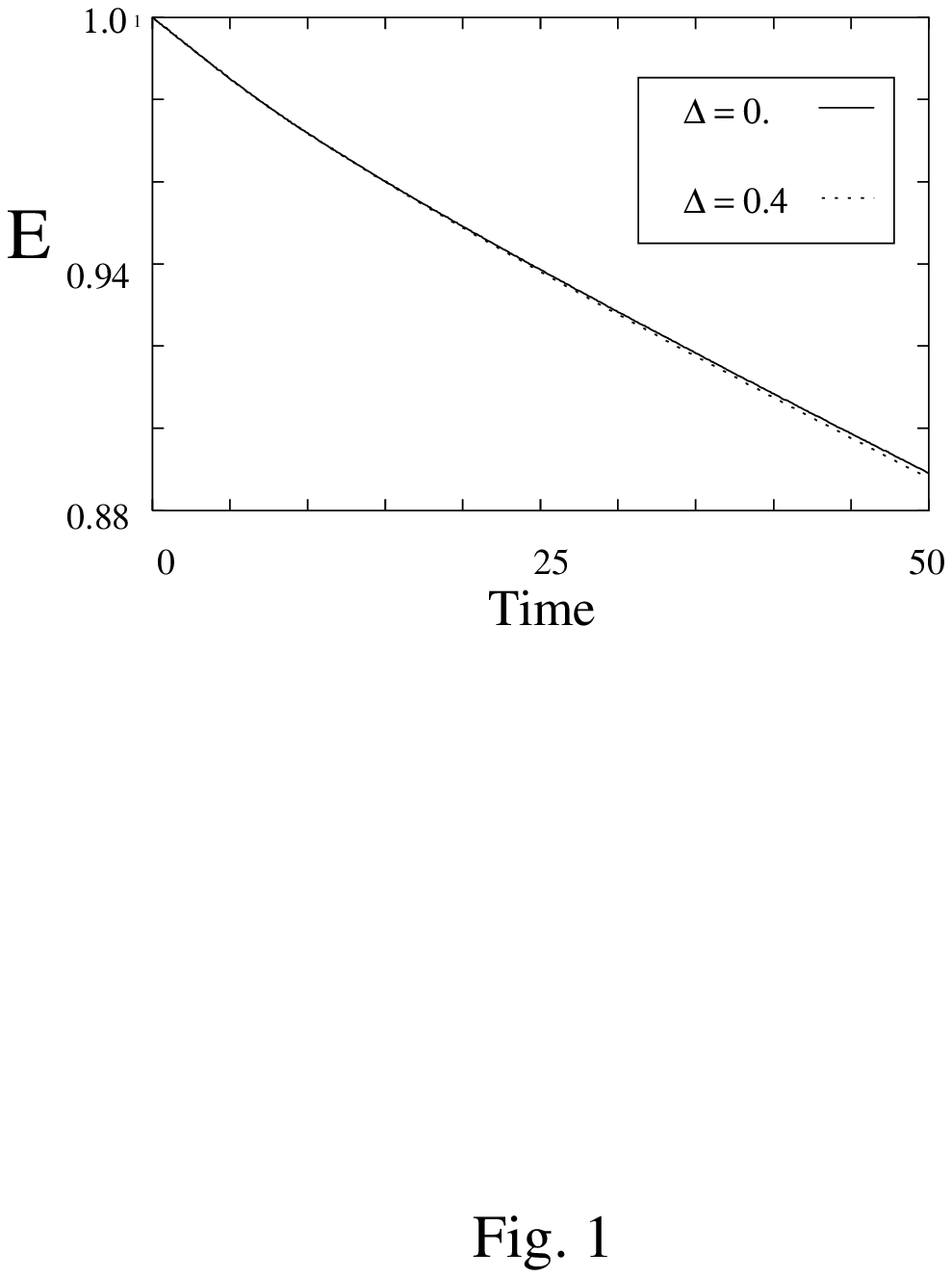,width=0.8\textwidth}}}
\end{figure}

\clearpage

\begin{figure}[p]
\centerline{\mbox{\psfig{file=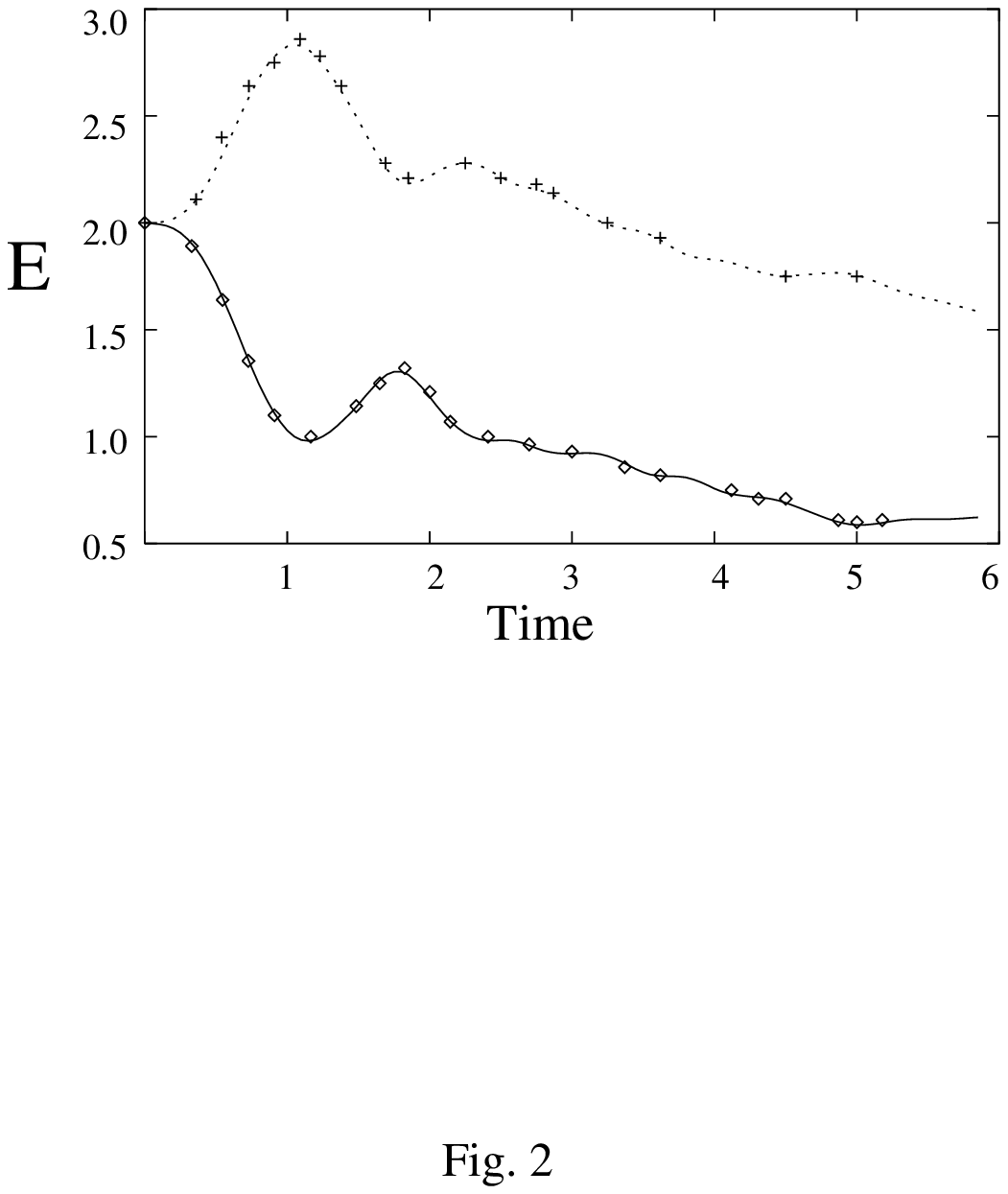,width=0.8\textwidth}}}
\end{figure}

\clearpage

\begin{figure}[p]
\centerline{\mbox{\psfig{file=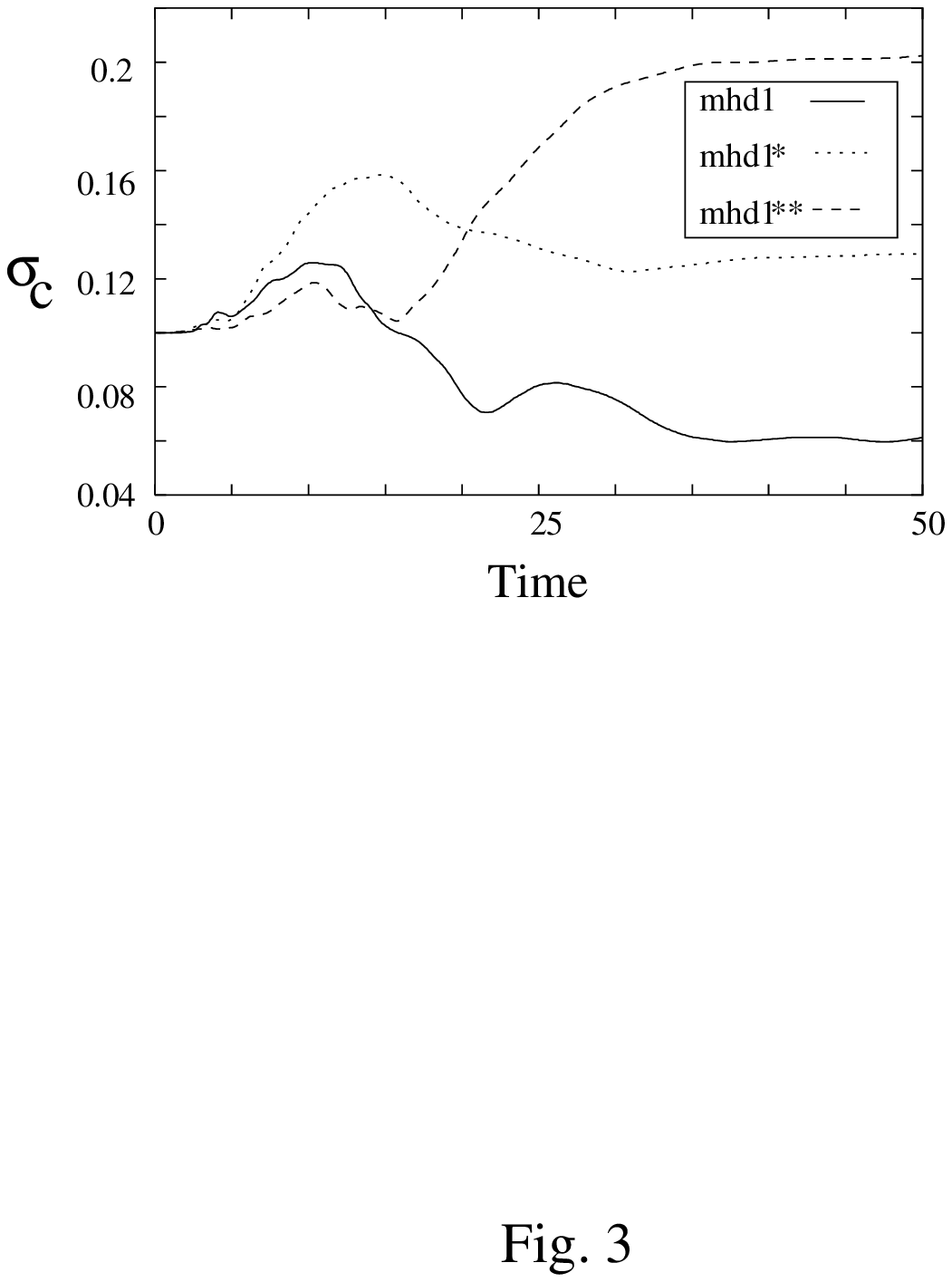,width=0.8\textwidth}}}
\end{figure}

\clearpage

\begin{figure}[p]
\centerline{\mbox{\psfig{file=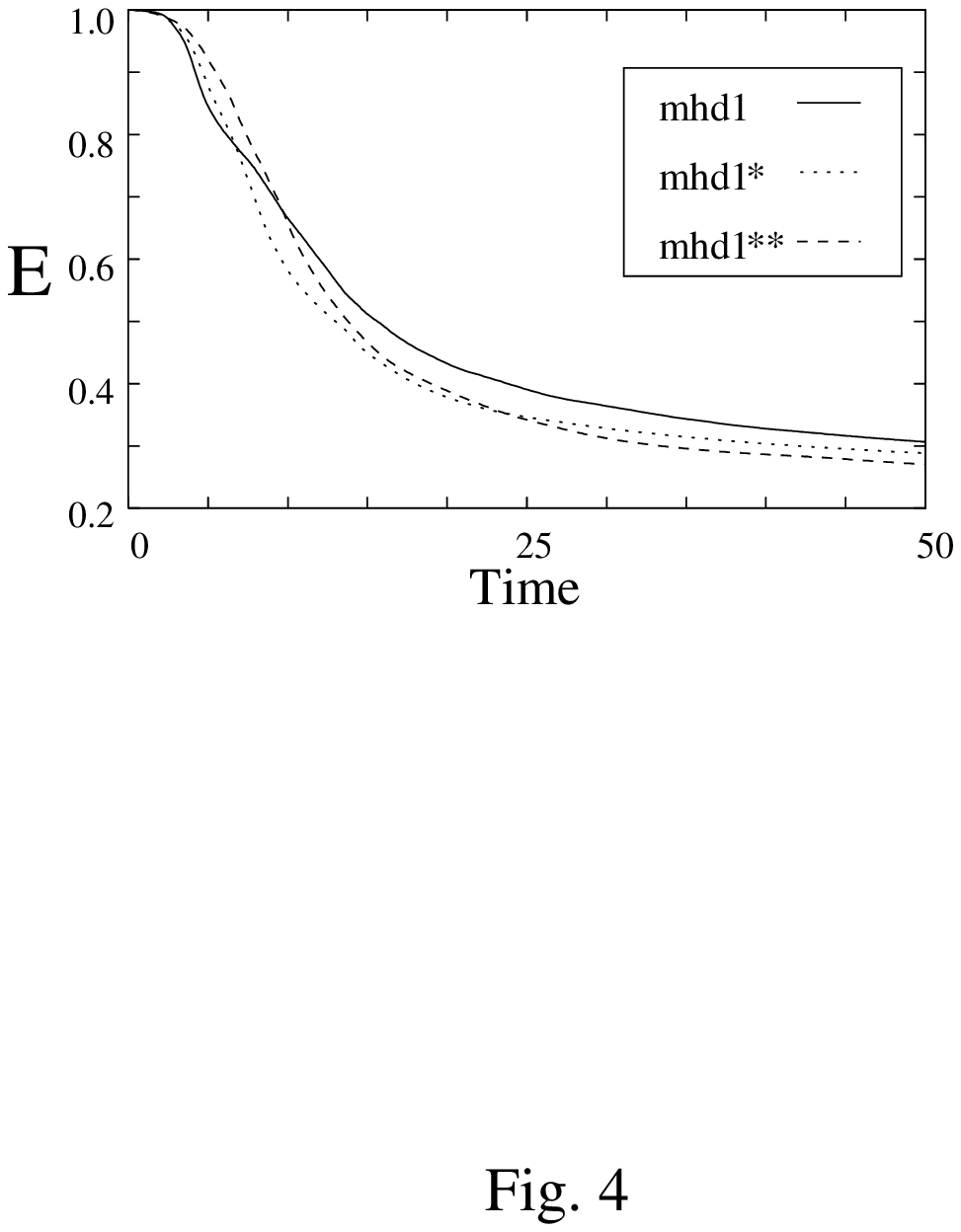,width=0.8\textwidth}}}
\end{figure}

\clearpage
\begin{figure}[p]
\centerline{\mbox{\psfig{file=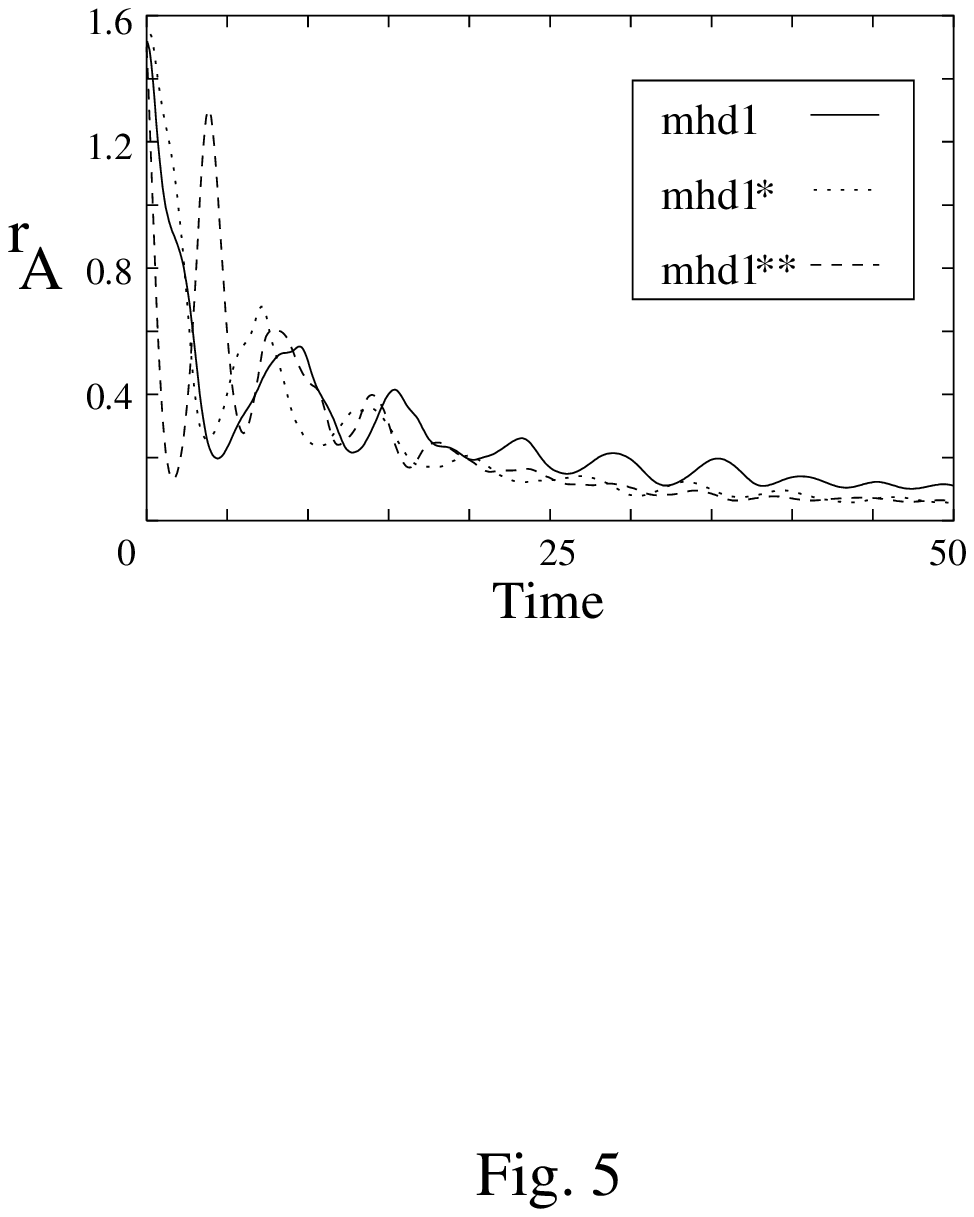,width=0.8\textwidth}}}
\end{figure}

\clearpage

\begin{figure}[p]
\centerline{\mbox{\psfig{file=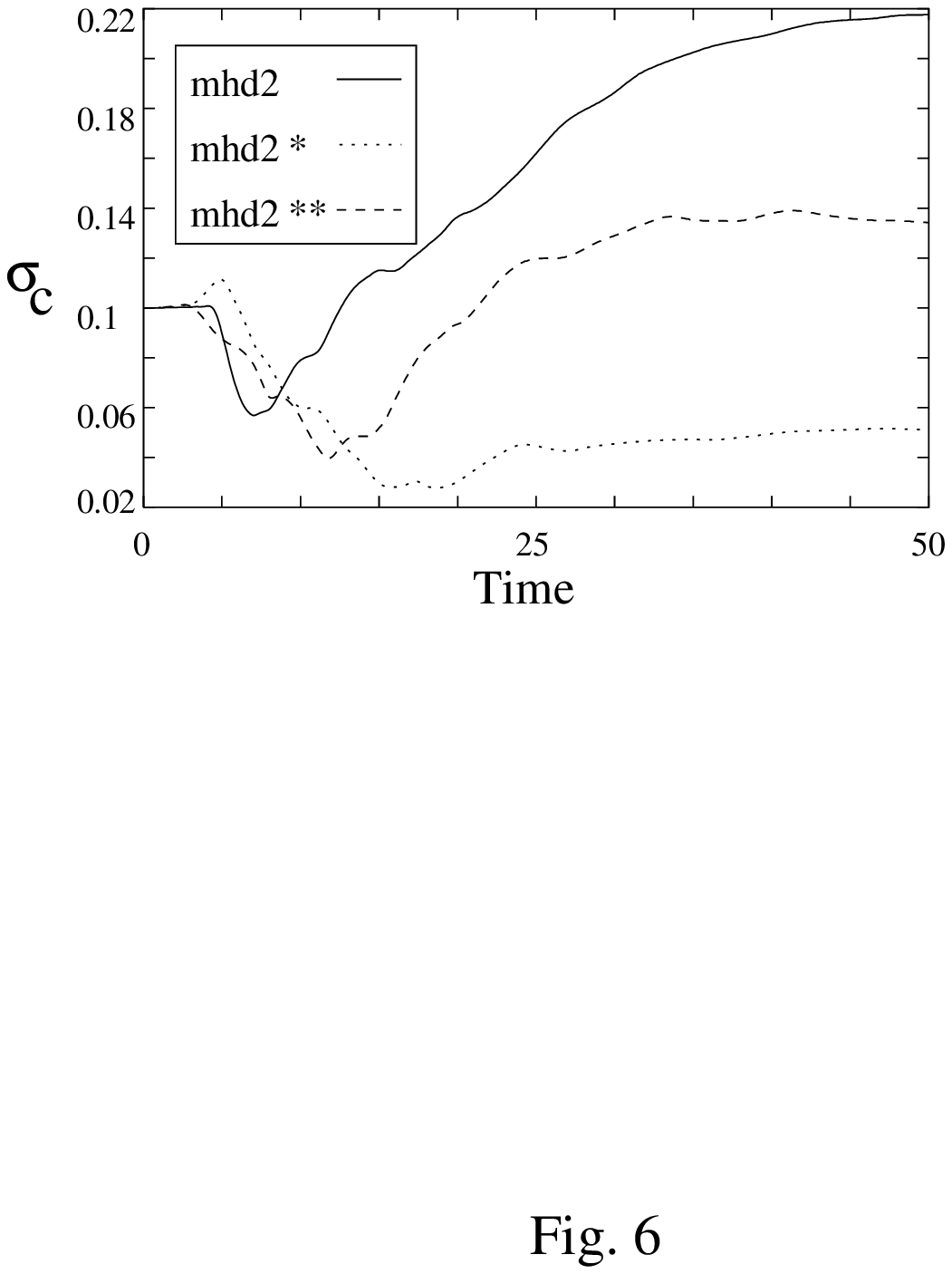,width=0.8\textwidth}}}
\end{figure}

\clearpage

\begin{figure}[p]
\centerline{\mbox{\psfig{file=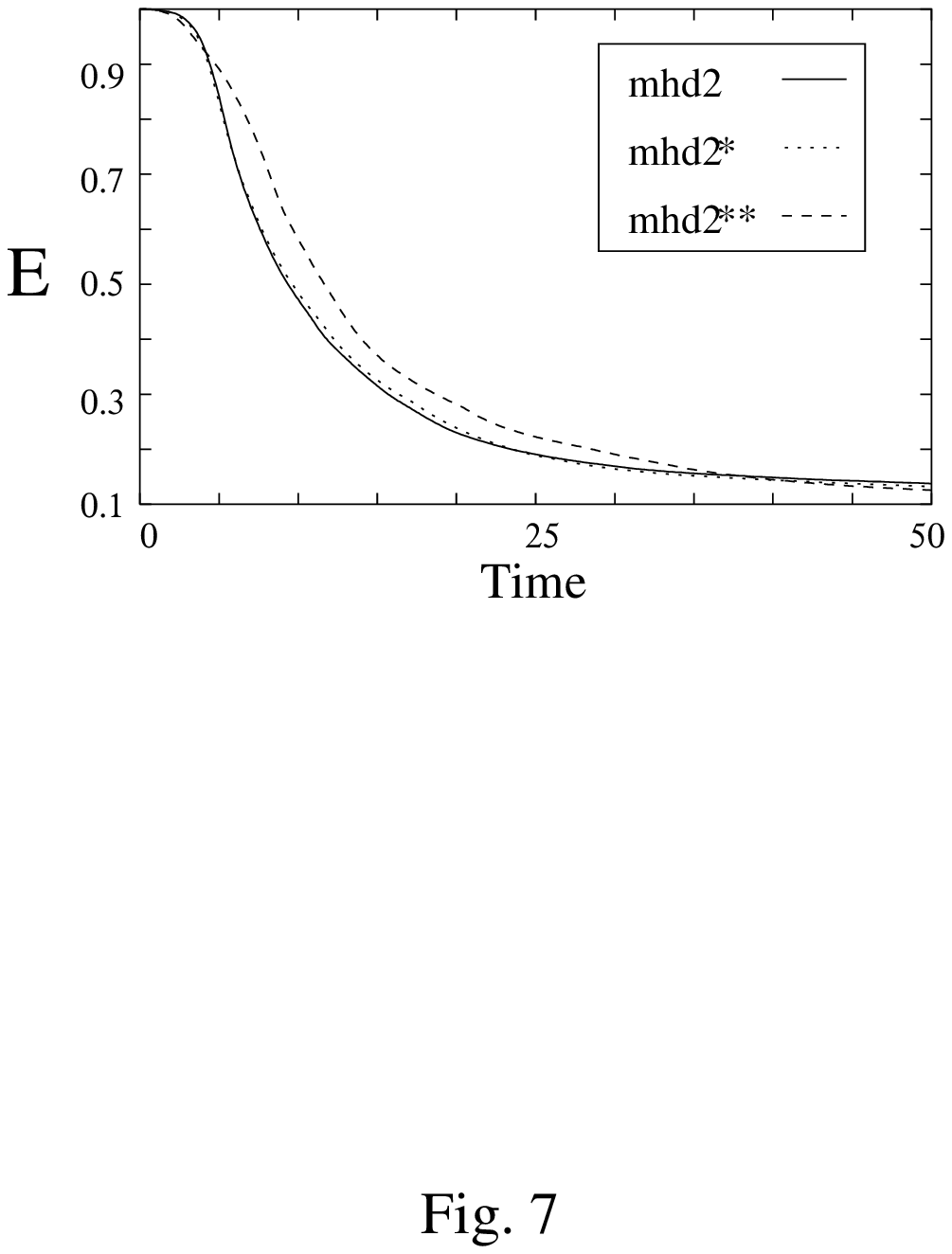,width=0.8\textwidth}}}
\end{figure}

\clearpage

\begin{figure}[p]
\centerline{\mbox{\psfig{file=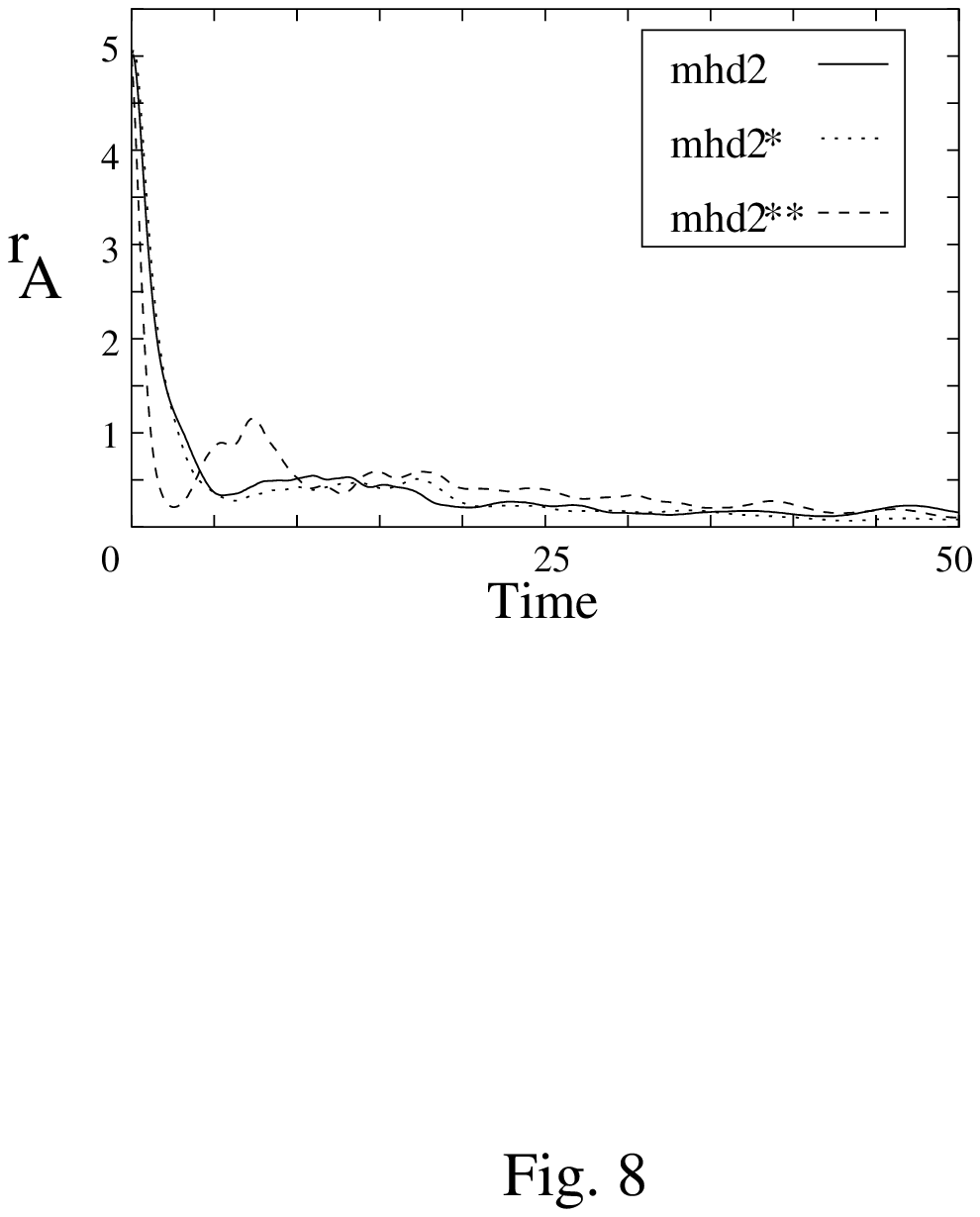,width=0.8\textwidth}}}
\end{figure}

\clearpage

\begin{figure}[p]
\centerline{\mbox{\psfig{file=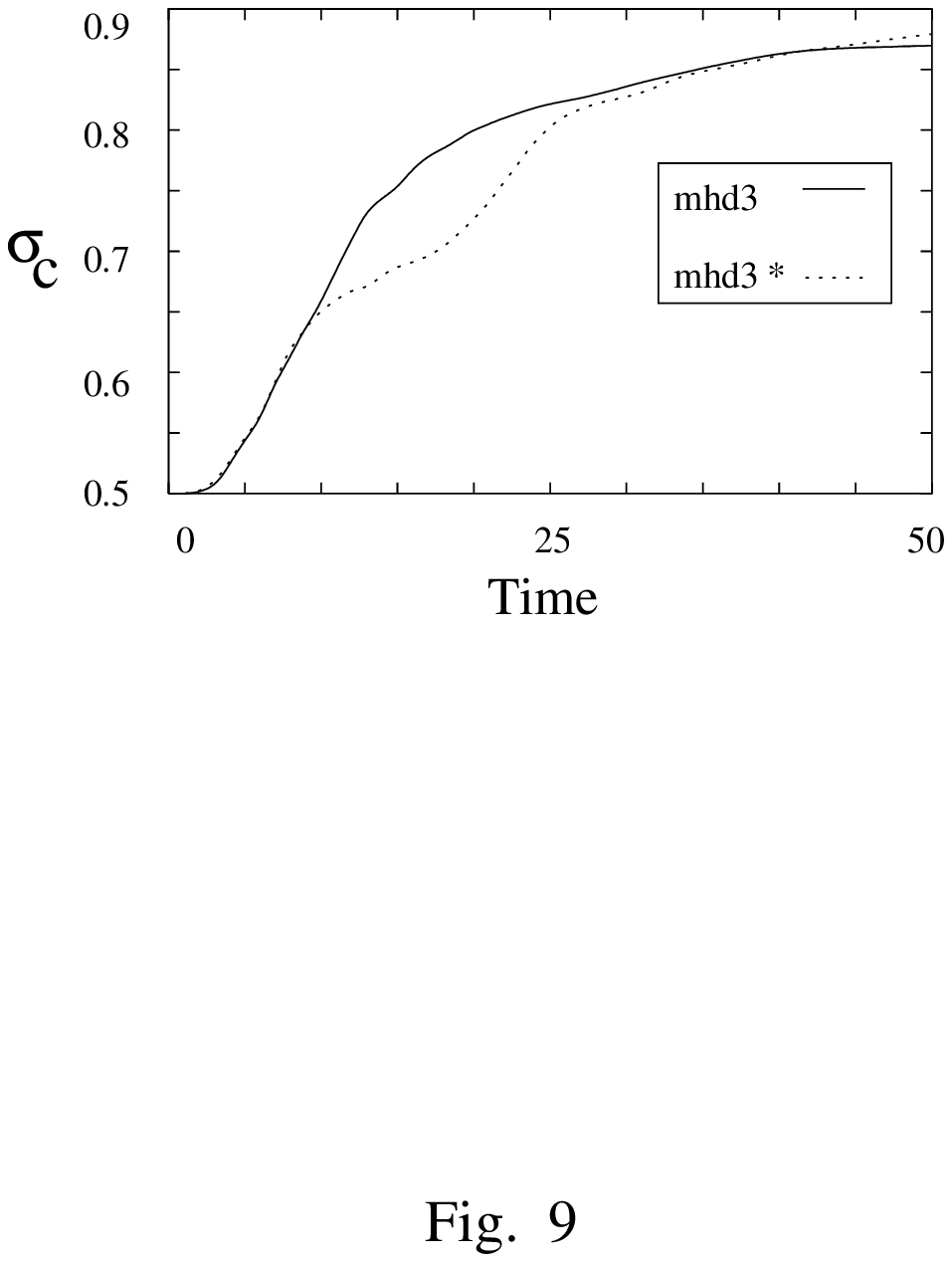,width=0.8\textwidth}}}
\end{figure}

\clearpage

\begin{figure}[p]
\centerline{\mbox{\psfig{file=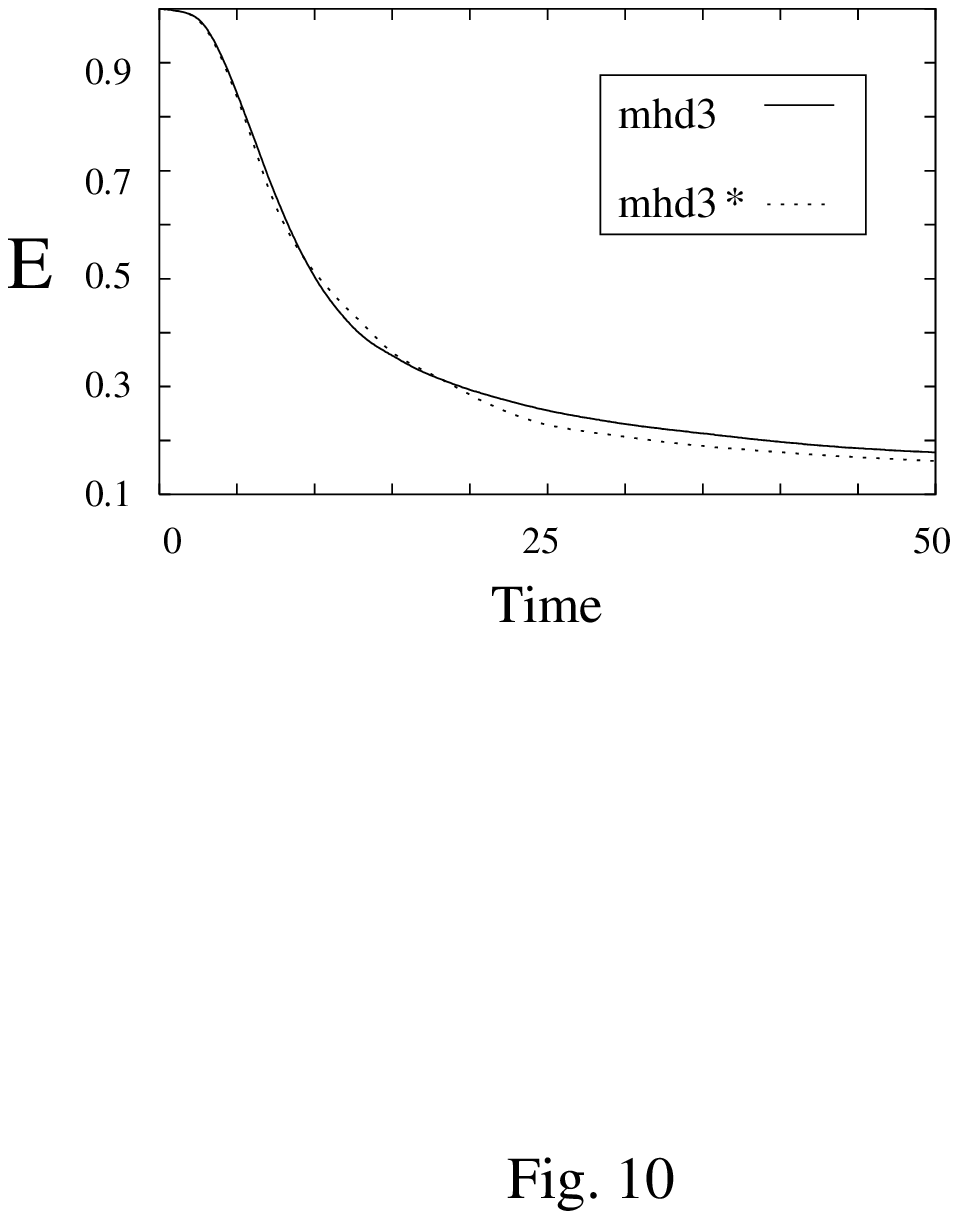,width=0.8\textwidth}}}
\end{figure}

\clearpage

\begin{figure}[p]
\centerline{\mbox{\psfig{file=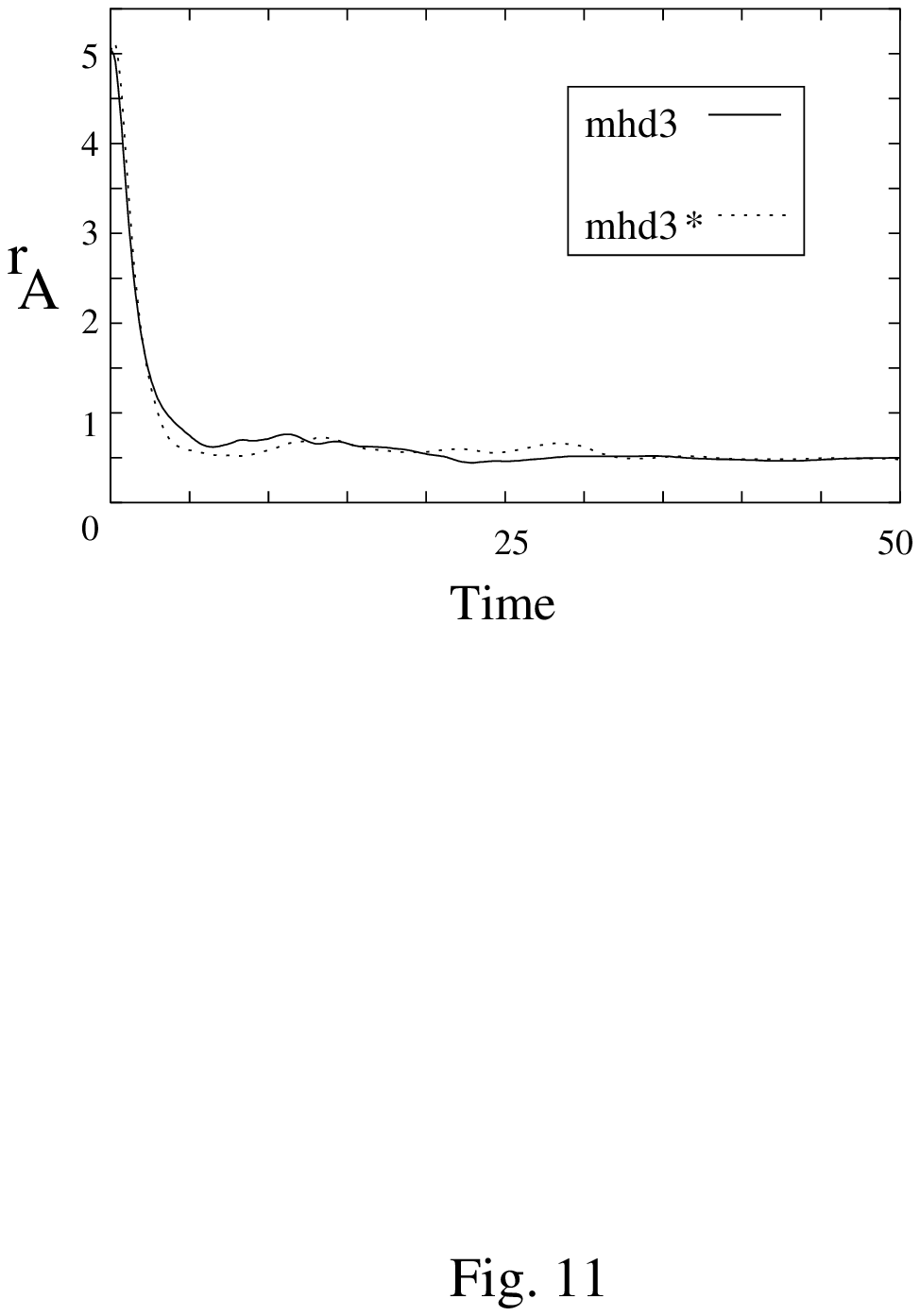,width=0.8\textwidth}}}
\end{figure}

\clearpage

\end{document}